\documentclass[useAMS,usenatbib,letters]{mn2e}

\usepackage{times}

\usepackage{amsfonts}
\usepackage{amssymb}
\usepackage{amsmath}
\usepackage{graphicx}

\def\apjl{ApJ}

\def\apss{Ap\&SS}
\def\aap{A\&A}

\def\mnras{MNRAS}

\def\rmxaa{Rev.~Mex.~Astron.~Astrofis.}

\def\p0{\phantom{0}}

\def\lessim{\raise-.5ex\hbox{$\buildrel<\over{\scriptstyle\mathtt{\sim}}$}}
\def\grtsim{\raise-.5ex\hbox{$\buildrel>\over{\scriptstyle\mathtt{\sim}}$}}

\title[Kinematics of Hen\,3-1333 and Hen\,2-113]{Spatially resolved kinematic observations of the planetary nebulae Hen\,3-1333 and Hen\,2-113\thanks{Based on observations made with the Australian National University 2.3\,m telescope at the Siding Spring Observatory under programme 1100147.
}} 
\author[A.~Danehkar and Q.\,A.~Parker]{A.~Danehkar$^{1,2}$\thanks{E-mail: ashkbiz.danehkar@students.mq.edu.au}
and
Q.\,A.~Parker$^{1,2,3}$ 
\\
$^{1}$Department of Physics and Astronomy, Macquarie University, Sydney, NSW 2109, Australia\\
$^{2}$Macquarie Research Centre in Astronomy, Astrophysics \& Astrophotonics, North Ryde, NSW 2109, Australia\\
$^{3}$Australian Astronomical Observatory, PO Box 915, North Ryde, NSW 1670, Australia
}

\begin{document}

\date{Accepted 2015 February 1.  Received 2015 January 30; in original form 2014 December 30}

\pagerange{L\pageref{firstpage}--L\pageref{lastpage}} \pubyear{2015}

\maketitle

\begin{abstract} 
We have performed integral field spectroscopy of the planetary nebulae Hen\,3-1333 (PNG332.9$-$09.9) and Hen\,2-113 (PNG321.0$+$03.9), which are unusual in exhibiting dual-dust chemistry and multipolar lobes but also ionized by late-type [WC\,10] central stars. The spatially resolved velocity distributions of the H$\alpha$ emission line were used to determine their primary orientations. The integrated H$\alpha$ emission profiles indicate that Hen\,3-1333 and Hen\,2-113 expand with velocities of $ \sim 32$ and $23$ km\,s$^{-1}$, respectively. The \textit{Hubble Space Telescope} observations suggest that these planetary nebulae have two pairs of tenuous lobes extending upwardly from their bright compact cores. From three-dimensional geometric models, the primary lobes of Hen\,3-1333 and Hen\,2-113 were found to have inclination angles of about $-30^{\circ}$ and $40^{\circ}$ relative to the line of sight, and position angles of $-15^{\circ}$ and $65^{\circ}$ measured east of north in the equatorial coordinate system, respectively.
\end{abstract}

\label{firstpage}


\section{Introduction}
\label{hen_3_1333:sec:introduction}

The planetary nebulae (PNe) Hen\,3-1333 (\,$=$\,PNG332.9$-$09.9) and Hen\,2-113 (\,$=$\,PNG321.0$+$03.9) show unique properties of the early post-asymptotic giant branch (AGB) phase and dust formation at the last AGB phase. \citet{Cohen1999,Cohen2002} found a dual-dust chemistry in these PNe using infrared observations. \citet{Cohen1999} fitted  blackbody curves to the continuum of Hen\,3-1333, and found some evidences for dual-chemistry, namely carbon-rich and oxygen-rich dust grains. Similarly, \citet{Cohen2002} identified oxygen- and carbon-rich materials in Hen\,2-113. 

The central stars of Hen\,3-1333 and Hen\,2-113 were among the coolest Wolf--Rayet stars studied in \textit{JHKL} bands and classified as [WC\,11] \citep{Webster1974}. They have been classified as late-type [WC10] based on the classification scheme proposed by \citet{Crowther1998}. \citet{Gleizes1989} derived the H~{\sc i} Zanstra temperature of $T_{\rm z}(\mbox{H\,{\sc i}})= 17.5$\,kK for Hen 3-1333 and $37$\,kK for Hen 2-113. Furthermore, photoionization modelling by \citet{DeMarco1998a} yielded $T_{\rm eff}= 25$\,kK for Hen\,3-1333 and $T_{\rm eff}= 29$\,kK for Hen\,2-113. 

\begin{figure}
\begin{center}
\includegraphics[width=2.5in]{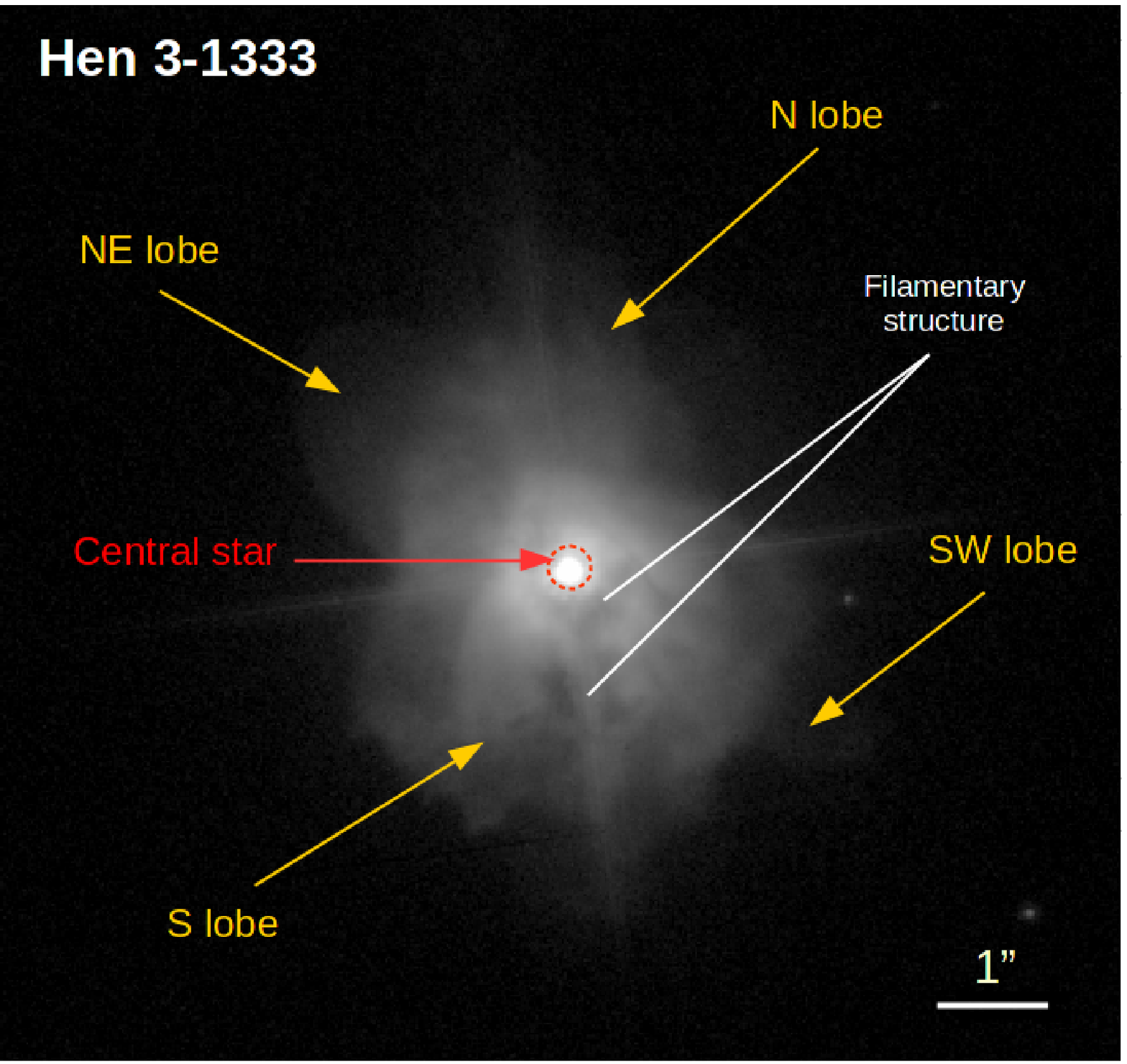}\\
\includegraphics[width=2.5in]{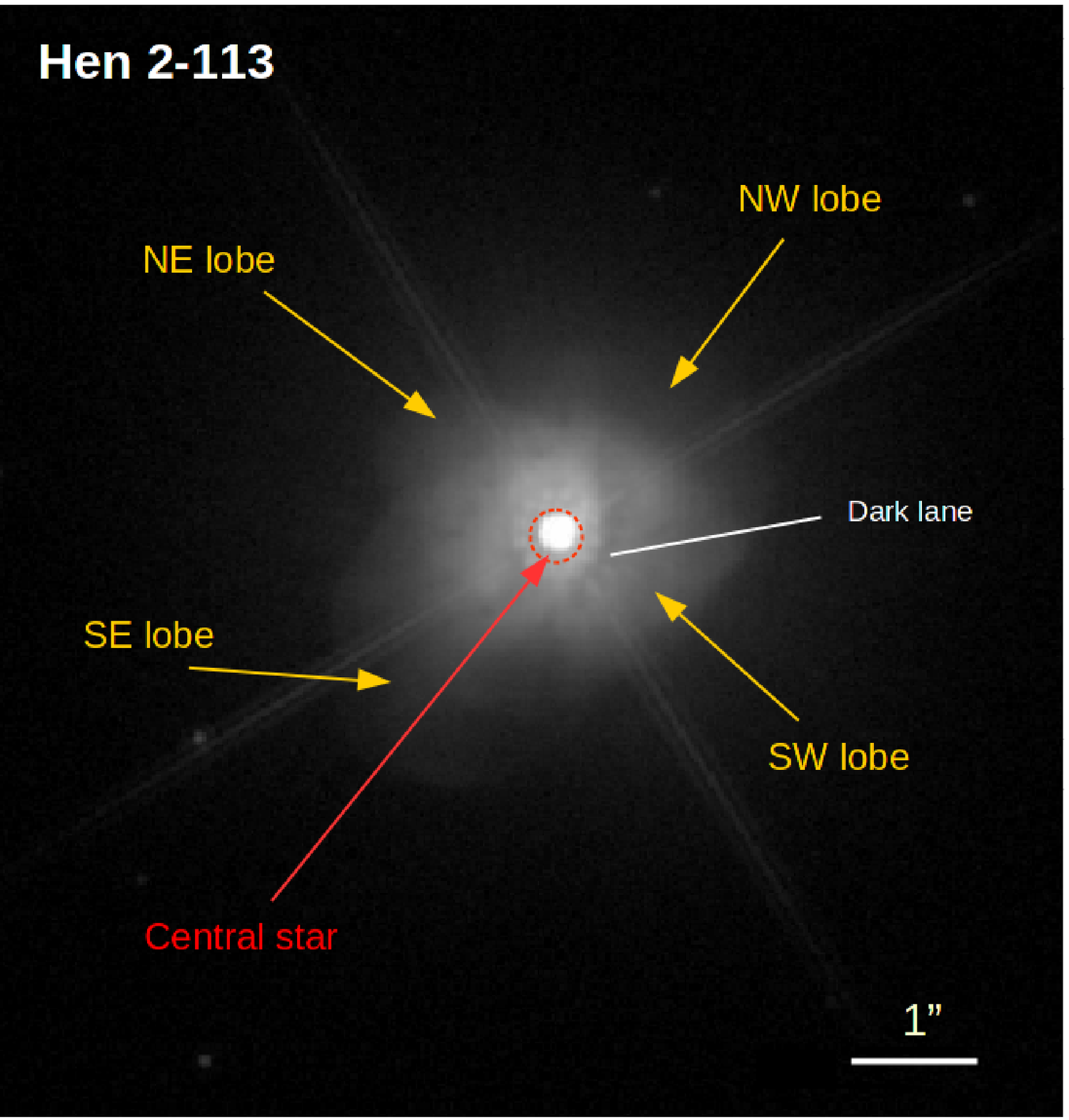}%
\caption{\textit{HST} images of Hen\,3-1333 (top panel) taken with the \textit{F}606\textit{W} filter, and Hen\,2-113 (bottom panel) taken with the \textit{F}814\textit{W} filter and ACS/HRC detector (Programme ID 9463, PI. R.~Sahai).
The image scale is shown by a solid line in each image. 
North is up and east is towards the left-hand side. 
\label{hen_3_1333:figures:hst}%
}%
\end{center}
\end{figure}

Fig.\,\ref{hen_3_1333:figures:hst} shows the \textit{Hubble Space Telescope (HST)} images of Hen\,3-1333 and Hen\,2-113 taken with the High Resolution Channel of the Advanced Camera for Surveys (ACS/HRC), and through the \textit{F}606\textit{W} and \textit{F}814\textit{W} filters, respectively. The \textit{HST} image of Hen\,3-1333 studied by \citet{Chesneau2006} hints at complex multipolar lobes surrounding its bright compact core. Previously, \citet{DeMarco2002} identified a compact dusty disk in Hen\,3-1333. Moreover, \citet{Lagadec2006} described the \textit{HST} image of Hen\,3-113 as two ring-like structures produced by the projection of a hourglass-shaped geometric model. 

In this paper, we present spatially resolved kinematic observations of Hen\,3-1333 and Hen\,2-113 made with an integral field unit (IFU) spectrograph. We first describe the observational method and the data obtained, and then proceed to determine the spatial orientations constrained by kinematic models.

\section{Observations}
\label{hen_3_1333:sec:observations}

The IFU observations were performed using the Wide Field Spectrograph \citep[WiFeS;][]{Dopita2007,Dopita2010} mounted on the 2.3-m Australian National University (ANU) telescope on 2010 April 20. Table \ref{hen_3_1333:tab:obs:journal} presents an observation journal, including the exposure time used for each PN in our WiFeS observations (column 3), and information on the \textit{HST} observations (columns 5--8).

\begin{table*}
\caption{Journal of observations.
\label{hen_3_1333:tab:obs:journal}
}
\centering
\footnotesize
\begin{tabular}{lllcccccc}
\hline\hline
Object & Other name & \multicolumn{2}{c}{WiFeS} & &\multicolumn{4}{c}{\textit{HST}} \\ 
\cline{3-4}\cline{6-9}
\noalign{\smallskip}
     &     &  Exp. (s)  &  Obs. Date & & Filter & Exp. (s)  &  Obs. Date &  Programme ID\\
\hline 
Hen\,3-1333 &PNG\,332.9$-$09.9 & 1200   & 2010 Apr 20 & & \textit{F}606\textit{W} & 56 & 2002 Sep 17 & 9463 \\ 
Hen\,2-113 &PNG\,321.0$+$03.9 & 60,1200 & 2010 Apr 20 & & \textit{F}814\textit{W} & 56 & 2003 Mar 09 & 9463 \\  
\hline
\end{tabular}
\end{table*}

WiFeS is an image-slicing IFU developed and built for the ANU, feeding a double-beam spectrograph. It has a field-of-view (FOV) of 25 arcsec $\times$ 38 arcsec  and a spatial resolution of  1 arcsec. We used the spectral resolution  of $R\sim 7000$. The classical data accumulation mode was used, so a suitable sky window has been selected from the science data for the sky subtraction purpose. We also acquired series of bias, dome flat-field frames, twilight sky flats, arc lamp exposures, wire frames and standard stars for bias-subtraction, flat-fielding and calibrations \citep[fully described by][]{Danehkar2014b}. 

The WiFeS data were reduced using the \textsc{iraf} pipeline {\sc wifes} (version 2.14). We utilized the same data reduction procedure described in detail by \citet{Danehkar2013} and \citet{Danehkar2014a}, which includes flat-fielding, bias-subtraction, wavelength calibration, spatial calibration and flux calibration. 

\begin{figure}
\begin{center}
{\footnotesize (a) Hen\,3-1333}\\ 
\includegraphics[width=1.60in]{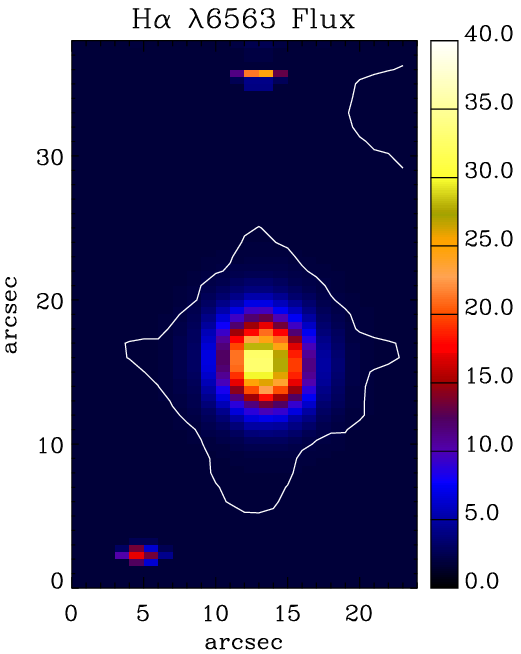}%
\includegraphics[width=1.60in]{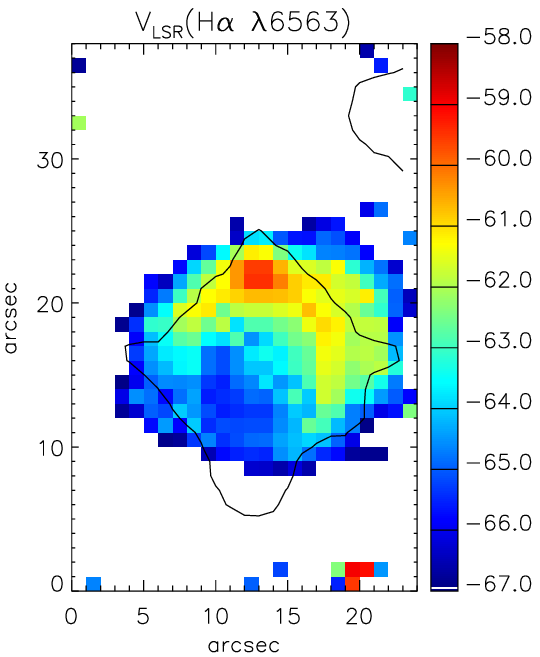}\\
{\footnotesize (b) Hen\,2-113}\\ 
\includegraphics[width=1.60in]{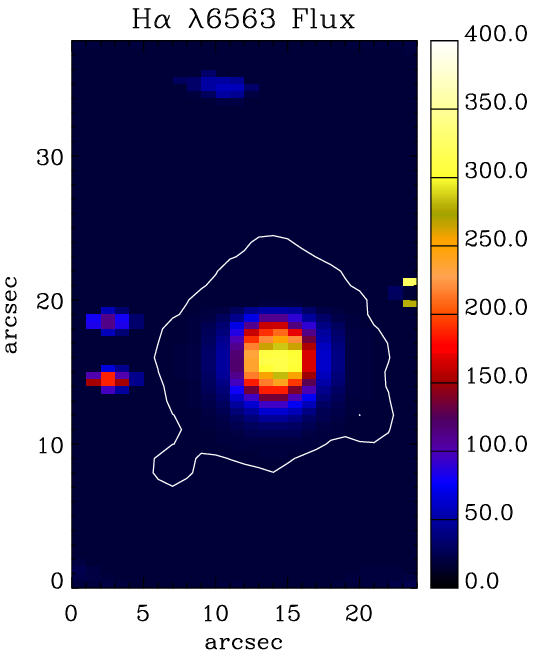}%
\includegraphics[width=1.60in]{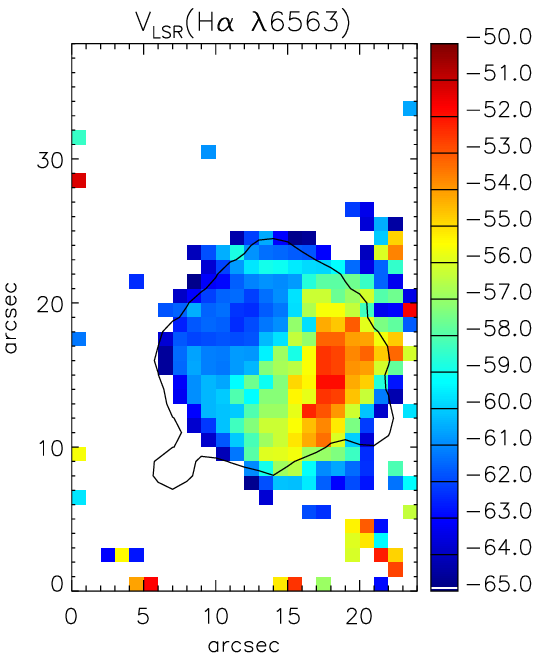}\\
\caption{From left to right, spatial distribution maps of flux intensity and LSR velocity of H$\alpha$ $\lambda$6563 for (a) Hen\,3-1333 and (b) Hen\,2-113. Flux unit is in $10^{-15}$~erg\,s${}^{-1}$\,cm${}^{-2}$\,spaxel${}^{-1}$ and velocities in km\,s${}^{-1}$. The white/black contour lines show the distribution of the narrow-band emission of H$\alpha$ in arbitrary unit obtained from the SuperCOSMOS H$\alpha$ Survey \citep{Parker2005}. North is up and east is towards the left-hand side. 
\label{fig:hen_3_1333}%
}%
\end{center}
\end{figure}

\section{Observational results}
\label{hen_3_1333:sec:results}

Fig.\,\ref{fig:hen_3_1333} presents spatially resolved maps of flux intensity and radial velocity derived from the H$\alpha$ $\lambda$6563 emission line for Hen\,3-1333 and Hen\,2-113. To extract them, we utilized a Gaussian curve fitting code, which allowed us to fit a single Gaussian profile to the H$\alpha$ emission line for all spaxels across the IFU field. As seen in Fig.\,\ref{fig:hen_3_1333}, the radial velocity maps indicate the presence of outer faint lobes with a diameter of about 10 arcsec extending to the compact structure of $\sim 3.5$ arcsec diameter (see Fig.\,\ref{hen_3_1333:figures:hst}). Contour lines in the figure depict the 2D distribution of the H$\alpha$ emission obtained from the SuperCOSMOS H$\alpha$ Sky Survey \citep{Parker2005}, which may aid us in distinguishing the nebular borders. However, the broader point spread function (PSF) of the stellar H$\alpha$ emission creates diffraction spikes around these PNe, so the contours do not actually show the nebular borders. The IFU flux map presents the resolved nebular H$\alpha$ emission.

\begin{table}
\footnotesize
\begin{center}
\caption{LSR systemic velocities and HWHM expansion velocities. 
}
\begin{tabular}{lcccccccl}
\hline\hline
Name 	& $v_{\rm sys}$(H$\alpha$) & \multicolumn{4}{c}{$V_{\rm HWHM}$(km\,s$^{-1}$)} \\
\cline{3-6}
\noalign{\smallskip}
       & (km\,s$^{-1}$) & H$\alpha$ &[N ~{\sc ii}] & [S ~{\sc ii}] & Mean \\
\hline
Hen\,3-1333& $-62.2$  & $31.6$  & $37.2$  & $30.6$ & $33.1\pm3.3$  \\
\noalign{\smallskip}
Hen\,2-113& $-56.7$   & $22.5$ & $22.3$ & $20.1$ & $ 21.6\pm1.2$ \\
\hline
\end{tabular}
\label{hen_3_1333:tab:kinematics1}
\end{center}
\end{table}

\begin{figure*}
\begin{center}
{\footnotesize (a) Hen\,3-1333}\\ 
\includegraphics[width=5.8in]{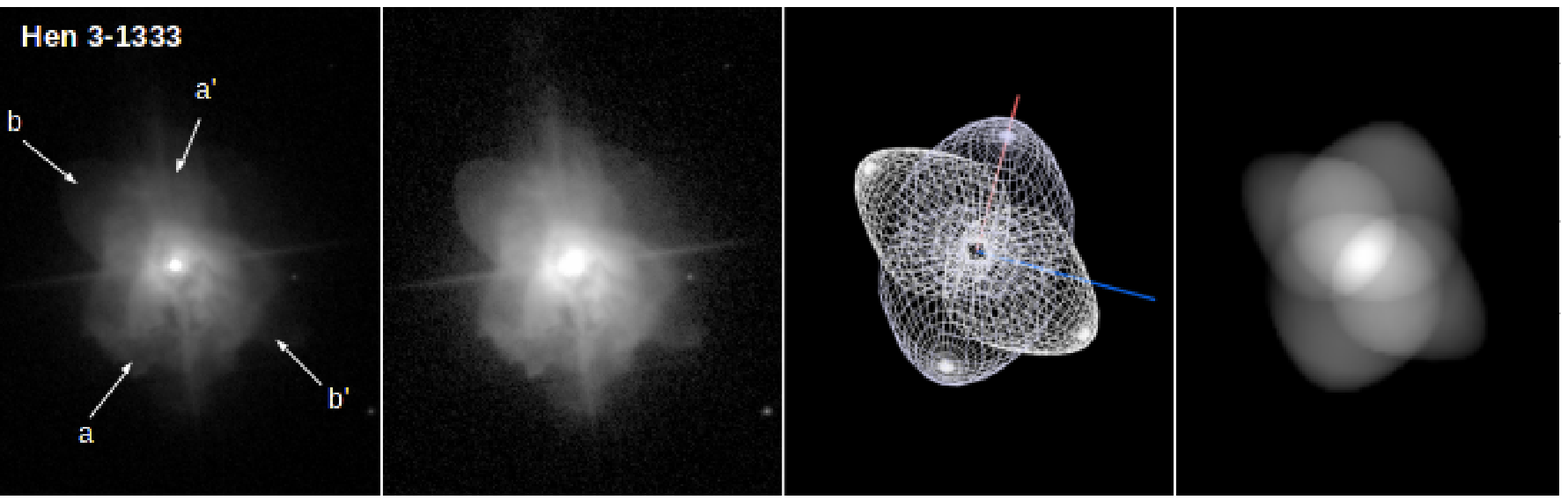}\\
{\footnotesize (b) Hen\,2-113}\\ 
\includegraphics[width=5.8in]{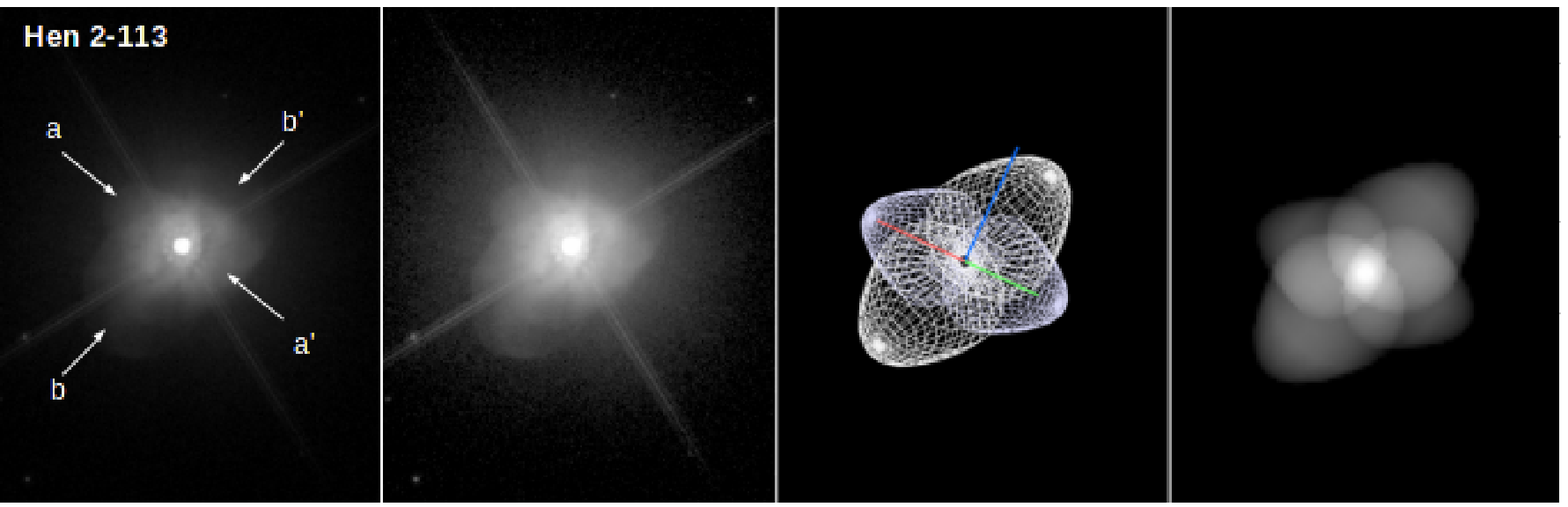}%
\caption{
\textit{HST} images at two different contrast levels (left-hand and second panels), \textsc{shape} mesh model (third panel), and rendered image (righ-handt panel) of (a) Hen\,3-1333 and (b) Hen\,2-113. The mesh models are shown at the best-fitting inclination. The synthetic rendered images have been obtained using a Gaussian blur.
\label{hen_3_1333:shap}%
}
\end{center}
\end{figure*}

Table~\ref{hen_3_1333:tab:kinematics1} lists our velocity results for different emission lines of the integrated spectrum of each PN. Column 2 presents the local standard of rest (LSR) systemic velocity ($v_{\rm sys}$) derived from the H$\alpha$ emission line.  Our expansion velocities ($v_{\rm exp}$) derived from the half width at half-maximum (HWHM) for H$\alpha$ $\lambda$6563, [N~{\sc ii}] $\lambda\lambda$6548,6584 and [S~{\sc ii}] $\lambda\lambda$6716,6731, and average HWHM values are presented in columns 3--6, respectively.  Our observations for the H$\alpha$ emission line give $V_{\rm HWHM}=31.6$ and $22.5$\,km\,s$^{-1}$ for Hen\,3-1333 and Hen\,3-113, in good agreement with $v_{\rm exp}=30$ and $19$\,km\,s$^{-1}$ derived by \citet{DeMarco1997}, respectively. For Hen\,3-113, \citet{Gesicki2006} also derived an expansion velocity of 18\,km\,s$^{-1}$ and a turbulence velocity of 15\,km\,s$^{-1}$.

From the \textit{HST} image of Hen\,3-1333 displayed in the top panel of Fig.\,\ref{hen_3_1333:figures:hst}, several elements can be identified: two pairs of bipolar lobes in the north and south, and some filamentary structures. The \textit{HST} image of Hen\,2-113 (shown in the bottom panel of Fig.\,\ref{hen_3_1333:figures:hst}), shows two pairs of bipolar lobes in the east and west, and a dark lane between the western border of two opposing lobes. But, the \textit{HST} images are also obscured by the PSF diffraction spikes, which need to be considered carefully when studying the morphology. 

\section{The models}
\label{hen_3_1333:sec:model}

\begin{table*}
\footnotesize
\begin{center}
\caption{The key parameters of the best-fitting kinematic models. }
\begin{tabular}{lcccccccccc}
\hline\hline
        &  & \multicolumn{4}{c}{Hen\,3-1333} &  & \multicolumn{4}{c}{Hen\,2-113}\\
\cline{3-6}\cline{8-11}
\noalign{\smallskip}
Feature &  & Size (arcsec)& PA($^{\circ}$)  & GPA($^{\circ}$)  & $i$($^{\circ}$)  &  & Size (arcsec) &PA($^{\circ}$)  & GPA($^{\circ}$)  & $i$($^{\circ}$) \\
\hline
$a-a'$ &  & $6.2 \times 3.7$ & $-15\pm5$ & $39.6\pm5$ & $-30\pm15$ &  & $3.7 \times 2.4$  & $65\pm5$ & $93.3\pm5$ & $40\pm15$ \\
\noalign{\smallskip}
$b-b'$ &  & $6.8 \times 2.9$ & $60\pm5$ & $114.6\pm5$ & -- &  & $4.8 \times 2.6$  & $120\pm5$ & $148.3\pm5$ & -- \\
\hline
\end{tabular}
\label{hen_3_1333:tab:shapemodel}
\end{center}
\end{table*}

The 3D kinematic modelling program \textsc{shape} \citep{Steffen2006,Steffen2011} was used to identify the spatial distribution of the outer tenuous lobes around the compact cores of Hen\,3-1333 and Hen\,2-113. The modelling procedure consists of defining a geometry, assigning emissivity distribution and defining a velocity law as a function of position. The program produces several outputs that can be directly compared with the observations, namely position--velocity diagram and appearance of the object on the sky. 

For this study, we adopted a bipolar geometric model (marked as $a - a'$ in Fig.\,\ref{hen_3_1333:shap}, left-hand panel). To replicate the \textit{HST} images, another bipolar component, which is identical to the other component, is introduced with different orientations (marked as $b - b'$ in Fig.\,\ref{hen_3_1333:shap}). We adjusted the inclination ($i$) and position angle `PA' in an iterative process until the qualitatively fitting solution is produced. The velocity IFU maps derived from the H$\alpha$ emission line, combined with the \textit{HST} images, were used to constrain the primary bipolar lobes ($a - a'$). The \textit{HST} observations were also used to reconstruct the secondary bipolar lobes ($b - b'$). Fig.\,\ref{hen_3_1333:shap} (second panel) also shows the \textit{HST} images with higher contrast, which reveal details not accessible at the previous contrast level.

Fig.\,\ref{hen_3_1333:shap} (third panel) shows a 3D representation of the model at the best-fitting inclination of each object. From the best-fitting kinematic models, the primary lobes ($a - a'$) of Hen\,3-1333 and Hen\,2-113 were found to have  PA~$=-15^{\circ}\pm5^{\circ}$ and $65^{\circ}\pm5^{\circ}$, and inclination angles of $i=-30^{\circ}\pm15^{\circ}$ and $40^{\circ}\pm15^{\circ}$, respectively. The secondary lobes ($b - b'$) of Hen\,3-1333 and Hen\,2-113, which are only noticeable in the \textit{HST} images, were found to have PA~$=60^{\circ}\pm5^{\circ}$ and $120^{\circ}\pm5^{\circ}$. Fig.\,\ref{hen_3_1333:shap} (right-hand panel) shows the synthetic rendered images. As seen, the rendered images perfectly resembles the \textit{HST} observations shown in Fig.\,\ref{hen_3_1333:shap} (left-hand panel). 

Table \ref{hen_3_1333:tab:shapemodel} lists the key parameters of the best-fitting morpho-kinematic models: the sizes, the PA, the Galactic position angle (GPA), and the inclination ($i$) of the lobes, respectively. The PA is the position angle of the bipolar lobes projected on to the plane of the sky, and measured from the north towards the east in the equatorial coordinate system  (ECS). The GPA is the position angle projected on to the sky plane, measured from the North Galactic Pole (NGP) towards the Galactic east. The inclination is measured between the line of sight and the nebular symmetry axis ($i=0^{\circ}$ being pole-on).

\section{Conclusion}
\label{hen_3_1333:sec:discussion}

The spatially resolved kinematic observations of Hen\,3-1333 and Hen\,2-113 presented in this paper have allowed us to identify their primary orientations. Our kinematic maps also indicate that these PNe have large extended faint lobes upwardly from their compact structures. For the primary lobes of Hen\,3-1333 and Hen\,2-113, we derived the inclination angles of $i=-30^{\circ}$ and $40^{\circ}$, respectively. Accordingly, the main orientations of Hen\,3-1333 and Hen\,2-113 were found to be PA~$=-15^{\circ}$ and $65^{\circ}$ in the ECS, in excellent agreement with the \textit{HST} studies by \citet{Chesneau2006} and \citet{Lagadec2006}, respectively. Interestingly, both Hen\,3-1333 and Hen\,2-113 have the same stellar characteristics \citep[][]{DeMarco1998a}, but they show different expansion velocities (see Table~\ref{hen_3_1333:tab:kinematics1}). Therefore, it seems that their nebular kinematic features are somehow unrelated to their stellar characteristics. 

Both Hen\,3-1333 and Hen\,2-113 demonstrate a dual-dust chemistry consisting of carbon-rich and oxygen-rich grains \citep{Cohen1999,Cohen2002,DeMarco2002}. \citet{Cohen2002} also found the same properties in other PNe with late-type [WC] stars. Moreover, \citet{Gorny2010} found more PNe with dual-dust chemistry in the Galactic bulge, and speculated that the simultaneous presence of O-rich and C-rich dust grains is more likely related to the stellar evolution in a close binary system. Recently, \citet{Guzman-Ramirez2014} identified the presence of dense central tori in PNe with dual-dust chemistry, suggesting the possible formation through a common-envelope phase. Meanwhile, some recent kinematic studies of bipolar PNe around close-binary central stars indicate that their binary orbital inclinations are very close to their nebular inclinations \citep[see e.g.][]{Jones2012,Tyndall2012,Huckvale2013}. Therefore, deeper observations of the central stars of Hen\,3-1333 and Hen\,2-113 will lead to a better understanding of their bipolar morphology and dual-dust chemistry. The presence of a stellar companion should be inspected. However, it is extremely difficult to detect a substellar companion, which could also affect the nebular properties.

\section*{Acknowledgements}

AD gratefully acknowledges the award of a Macquarie Research Excellence Scholarship (MQRES). QAP acknowledges support from Macquarie University and the Australian Astronomical Observatory (AAO). We would like to thank the
staff at the ANU Siding Spring Observatory for their support. Based on observations made with the NASA/ESA \textit{Hubble Space Telescope}, and obtained from the Hubble Legacy Archive, which is a collaboration between the Space Telescope Science Institute (STScI/NASA), the Space Telescope European Coordinating Facility (ST-ECF/ESA) and the Canadian Astronomy Data Centre (CADC/NRC/CSA). We thank the anonymous referee for helpful comments.

\end{document}